\documentclass[12pt]{article}

\usepackage[dvips]{graphicx}

\def\dspace{\baselineskip = 0.30in}

\def\lapproxeq{\lower .7ex\hbox{$\;\stackrel{\textstyle
<}{\sim}\;$}}
\def\gapproxeq{\lower .7ex\hbox{$\;\stackrel{\textstyle
>}{\sim}\;$}}

\begin{document}

\dspace

\begin{titlepage}
\begin{flushright}
%\preprint{
BA-02-08\\
%March 2002\\
%}
\end{flushright}
\vskip 2cm
\begin{center}
%\title
{\Large\bf Non-universal
Soft Parameters in Brane World
and the Flavor Problem in Supergravity}
\vskip 1cm
{\normalsize\bf
%\author{
Bumseok Kyae\footnote{bkyae@bartol.udel.edu} and
Qaisar Shafi\footnote{shafi@bxclu.bartol.udel.edu}}
\vskip 0.5cm
{\it Bartol Research Institute, University of Delaware, \\Newark,
DE~~19716,~~USA\\[0.1truecm]}

%
%%\maketitle

\end{center}
\vskip .5cm

%\date{\today}
%\pacs{PACS: 11.25.Mj, 12.10.Dm, 98.80.Cq}

\begin{abstract}

We consider gravity mediated supersymmetry (SUSY) breaking in 5D spacetime
with two 4D branes B1 and B2 separated in the extra dimension.
Using an off-shell 5D supergravity (SUGRA) formalism, we argue that the
SUSY breaking scales could be non-universal even at the fundamental scale
in a brane world setting, since SUSY breaking effects could be effectively
localized.  As an application, we suggest a model in which
the two light chiral MSSM generations reside on B1, while the third generation
is located on B2, and the Higgs multiplets as well as gravity and
gauge multiplets reside in the bulk.
For SUSY breaking of the order of 10--20 TeV caused
by a hidden sector localized at B1,
the scalars belonging to the first two generations can become sufficiently
heavy to overcome the SUSY flavor problem.
SUSY breaking on B2 from a different localized
hidden sector gives rise to the third generation soft scalar masses
of the order of 1 TeV.  Gaugino masses are also of the order of 1 TeV
if the size of the extra dimension is $\sim 10^{-16}$ ${\rm GeV}^{-1}$.
As in 4D effective supersymmetric theory,
an adjustment of TeV scale parameters
is needed to realize the 100 GeV electroweak symmetry breaking scale.

\end{abstract}
\end{titlepage}

\newpage

%%%%%%%%%
%%%%%%%%%

%%%%%%%%%%%%%%%%%%%%%%%%%%%%%%%%%%%%%%%%%%%%%%%%%%%%%%%%%%%%%%%%%%%%%%%%%%%
\section{Introduction}

The minimal supersymmetric standard model (MSSM) is an attractive avenue
for physics beyond the standard model (SM).
The MSSM not only resolves the naturalness problem ~\cite{witten}
regarding the small Higgs scalar mass that SM suffers from, but also
improves the scenario of gauge couplings unification~\cite{unif} and
electroweak symmetry breaking~\cite{smbreak}.
However,
the MSSM has numerous ($>$ 100) theoretically undetermined parameters.
To understand the origin of many soft supersymmetry (SUSY) breaking
parameters in the MSSM simply and consistently,
one often invokes the supergravity (SUGRA) models.
However, models relying on the conventional gravity-mediated SUSY
breaking scenario~\cite{nilles} generally contain flavor changing processes
that are strongly constrained by experiments~\cite{experiment}.
Approaches for resolving this notorious problem include
gauge-mediated SUSY breaking~\cite{gaugemedi} and
anomaly mediated SUSY breaking~\cite{anomalymedi}.
On the other hand the effective supersymmetric theory
(ESUSY)~\cite{esusy,u1a,radia} provides an alternative scenario.
In ESUSY, the superpartners of the first two generations are required to
be sufficiently heavy ($\sim$ 20 TeV) to suppress flavor changing processes
in SUGRA models.
(This also could provide a mechanism
for suppressing dimension five proton decay operators.)
Because of the relatively small Yukawa couplings of the first two generations,
the Higgs boson masses are radiatively stable despite these heavy masses.
On the other hand, the left- and right-handed top squarks are constrained
to be not much heavier than 1 TeV or so
to preserve the gauge hierarchy solution
because of the relatively large top quark Yukawa coupling~\cite{smbreak}.
One expects from $SU(2)$ symmetry that the left-handed sbottom mass
also does not exceed a TeV or so.
In addition, the B-ino, W-ino, and also Higgsino, which
are coupled to the Higgs scalars with ``sizable'' gauge couplings,
should remain lighter than a TeV or so.\footnote{In Ref. \cite{u1a},
an ESUSY idea is realized by introducing an anomalous $U(1)_A$. }

However, with only a unique SUSY breaking scale at $M_{\rm Planck}$ as in
ordinary 4D SUGRA models, such a large hierarchy between the first two and
the third generations' soft masses is not easily derived radiativlely
at the electroweak scale, even though the first two generations' Yukawa
couplings are relatively small.\footnote{Reference \cite{radia} shows
that the 10 TeV--sub-TeV hierarchy
between the first two and the third generations soft scalar masses at
the electroweak scale can be derived radiatively from 10 TeV soft masses at
the GUT scale.}
Moreover, the first two generations' heavy soft mass squareds tend to drive
the third ones to negative values at low energy
through two loop renomalization group equations,
so that the third generation squark
masses should be raised to several TeV at the grand unified theory (GUT)
scale \cite{arkani}.
This makes the scenario perhaps somewhat less attractive.

In this paper we will introduce two (or more) SUSY breaking scales
at the GUT or some other fundamental scale within a gravity-mediated SUSY
breaking scenario without invoking any flavor symmetry,
deploying two (or more) 4 dimensional branes or orbifold fixed points
in 5 dimensional spacetime.
In 4D gravity-mediated SUGRA models
using only hidden sector SUSY breaking,
there is no ``easy'' way to provide several SUSY breaking effects
that depend on the flavor;
once SUSY breakes down in a hidden sector,
the consequences typically spread impartially
among particles in the visible sector
through direct gravity couplings appearing in the SUGRA Lagrangian.
However, if we could somehow couple each visible sector field
{\it gravitationally} exclusively to a specific hidden sector
among several hidden sectors
with different SUSY breaking scales,
it would be possible to introduce flavor dependent
SUSY breaking scales even within the gravity-mediated scenario.
This is where the extra dimension and branes can play an important role.

The SUSY breaking soft scalar masses and the ``$A$'' terms are derived
from the scalar potential in SUGRA models.
In 5D brane world, the scalar potentials as well as the Yukawa interaction
terms are
usually constructed only on the branes due to supersymmetry.
Thus, as shown in \cite{peskin} for global SUSY,
the scalar potentials are localized by the delta function on the
brane and separated from each other at tree level.
In this paper we will show that the SUSY breaking effects
from the hidden sectors localized on the branes
are transmitted to the visible sector fields located
on the same brane and bulk gauginos
through direct gravitational couplings, but
remain shielded at least at tree level
from the other brane fields~\cite{anomalymedi,dine}
and bulk scalars.
Thus if the SUSY breaking scales are $m_1$ and $m_2$
at two separated branes B1 and B2, respectively,
the localized chiral multiplets at B1 (B2) could get soft SUSY breaking effects
of order $m_1$ ($m_2$).
The mixing terms from one loop contribution among fields from two distinct
branes can be sufficiently suppressed
if the interval size is sufficiently large compared to the fundamental
scale~\cite{gherg}.

Let us assume that the SUSY breaking scales at B1 and B2 are
$m_1$ (10--20 TeV) $>$ $m_2$ ($\sim$ 1 TeV).
Then, we can easily make the superpartners of the first two generations
sufficiently heavy by locating them on B1.
In order to keep the radiative corrections to the Higgs masses under control,
the third family resides on B2 (or in the bulk).
The two Higgs multiplets should reside in the bulk, so that
the first two and the third generations of the quarks and leptons can couple
to them at B1 and B2, respectively.
Note that matter fields in the bulk are accompanied by bulk gauge fields
for the consistency of the gauge theory.
%According to the above discussion,
%bulk gauginos would possibly acquire $O$(10 TeV) soft masses from B1.

The gaugino mass term can be generated at tree level
in some D=5 off-shell SUGRA formalism~\cite{zucker2}
if SUSY is spontaneously broken,
even when a singlet field does not couple to the gauge kinetic term.
Thus heavy gaugino masses from SUSY breaking at B1
could give rise to large radiative corrections
to the Higgs mass and spoil the naturalness solution.
However, since the bulk gaugino mass is given
by $m_{1/2}\sim m_{3/2}\times (M_*/M_P)$,
where $M_*$ is a 5D SUGRA fundamental scale and
$M_P$ is the reduced 4D Planck scale,
a not so ``large'' extra dimension,
for example, of order $~10^{-16}$ ${\rm GeV^{-1}}$
(or $M_*\sim 10^{17}$ GeV), suppresses the gaugino masses
to 1 TeV scale.
Moreover, by introducing an additional set of the localized gauge multiplet
at B2 with large coefficient ($\sim$ 10)
(i.e., in addition to the bulk gauge multiplet),
the gauginos mass can be more suppressed to the 100 GeV scale
as we will see.
In such a way, the masses of the various gauginos in the MSSM
also could be made non-universal.

%
%In our paper, we only consider hidden sectors that are localized on the branes
%and simply assume suitable vacuum expectation value (VEV) of the hidden sector
%superpotential.
%However, our arguments could be extended to more general cases
%like gaugino condensation.
%Even if we would introduce a unique hidden sector in the bulk,
%the hidden sector gauge
%couplings at each branes and bulk could be different, only if we introduce
%additional localized gauge muliplets also.
%Then, hidden sector gaugino could condense only on the branes
%even with only one
%hidden sector gauge group,
%if its gauge coupling is large enough only on the corresponding branes by
%the additinal brane gauge multiplets.

\section{5D off-shell SUGRA and 4D brane matter}

In order to couple 5D SUGRA to 4D brane matter
it is convenient to employ the 5D off-shell SUGRA
formalisms recently developed in~\cite{zucker1,zucker3,zucker2,zuckerth,kugo}.
Although we will employ the formalism
in~\cite{zucker1,zucker3,zucker2,zuckerth},
our essential results such as localized SUSY breaking effects
and realization of ESUSY should also hold in other formalisms and/or in other
spacetime dimensions.

The 5D formalism in \cite{zucker1} is an extension of the 4D SUGRA off-shell
formalism of~\cite{sohnius}.
The latter contains (16+16) bosonic and fermionic
degree of freedoms, namely, $e^a_m$ (vierbein), $\psi_m$ (gravitino),
$a_m$ ($U(1)_R$ axial gauge vector), $b_a$ (axial vector),
$\xi$ (spinor), $S$ (scalar), $t^1$ (pseudoscalar), and $t^2$ (scalar),
where every field is auxiliary except for the vierbein and the gravitino.
In 5D, N=2 SUGRA, the familiar N=2 SUGRA on-shell
fields, namely the f${\rm \ddot{u}}$nfbein ($e^A_M$), gravitino ($\psi^j_M$)
($j=1,2$)
and graviphoton ($A_M$) are supplemented by the auxiliary fields,
\begin{eqnarray}
\vec{t} ~, ~~~v_{AB} ~, ~~~
\vec{V}_M ~, ~~~
\zeta ~,~~{\rm and}~~~  C ~,
\end{eqnarray}
an isotriplet, an antisymmetric tensor, a $SU(2)_R$ gauge vector,
a spinor, and a scalar, respectively.
In the gravitino $\psi_M^j$, the index ``$j$'' (=1,2) is
a $SU(2)_R$ index.
Although the closure of SUSY algebra only requires
the above (40+40) bosonic and fermionic degrees of freedom
(``minimal multiplet''),
the corresponding action turns out to be unphysical; the equations of
motion give rise to ${\rm det}~e^A_M=0$.
(This problem is also observed in 4D SUGRA formalism
with a single auxiliary fermion \cite{sohnius}.)
To resolve it, additional (8+8) auxiliary fields are needed, either from
a nonlinear multiplet (version I) \cite{zucker1},
a hypermultiplet (version II) \cite{zuckerth}, or
a tensor multiplet (versin III) \cite{zucker3}.
In this paper we prefer the first choice.
The nonlinear multiplet contains the auxiliary fields,
$\phi^j\,_\alpha$ (scalar), $~\chi$ (spinor), $~\varphi$ (scalar),
and $~V_A$ (vector),
where $\alpha$ is the index of an additional $SU(2)$ and $a$ is the Lorentz
index.
Closure of the algebra on $~\varphi$ and $~V_A$ gives rise to
the constraint \cite{zucker1}
\begin{eqnarray} \label{constraint}
C-\frac{1}{16\kappa}R_{AB}\,^{AB}-\frac{\kappa}{24}F_{AB}F^{AB}
+5\kappa\vec{t}^2+\frac{\kappa}{4}v_{AB}v^{AB}
-\frac{1}{8\kappa}\hat{\cal D}_M\phi^j{}_\alpha\hat{\cal D}^M\phi^\alpha{}_j
+\cdots =0 ~,
\end{eqnarray}
where $\hat{\cal D}_M$ is the supercovariant derivative
defined in \cite{zucker1},
and ``$\cdots$'' contains the fermionic contributions.
The bosonic part of the pure SUGRA action is given by \cite{zucker1}
\begin{eqnarray}\label{sugra}
S_{SUG}&=&\int d^4x\int_{-L}^{L}dy~
e_5\frac{1}{2L}\bigg[\frac{-1}{4\kappa^2}R(\hat{\omega})_{AB}{}^{AB}
-\frac{1}{6}F_{AB}F^{AB}
-\frac{1}{\sqrt{3}}F_{AB}v^{AB}
\nonumber \\
&&-\frac{\kappa}{6\sqrt{3}}\epsilon^{ABCDE}A_AF_{BC}F_{DE}
%-\frac{1}{\sqrt{3}}F_{AB}v^{AB}
+v_{AB}v^{AB}
-12\vec{t}^2 -\frac{1}{4}\vec{V}_A\vec{V}^A  \\
&&-\frac{1}{2\kappa^2}\hat{\cal D}_M\phi^i{}_\alpha\hat{\cal D}^M\phi^\alpha{}_i
+V_AV^A-\varphi^2 ~\bigg]  ~, \nonumber
\end{eqnarray}
where the factor $2L$ for the size of the extra dimension ensures that the
mass dimensions of all fields correspond to their canonical 4D ones,
and the induced SUGRA parameter $\kappa$ in Eq.~(\ref{sugra}) and the original
5D SUGRA parameter $\kappa_*$ are defined as
\begin{eqnarray}
\kappa^2&\equiv& 8\pi G\equiv \frac{1}{M_P^2} ~, \\
\label{fundamental}
\kappa_*^3&\equiv& \frac{1}{M_*^3}\equiv 2L\times\kappa^2 =\frac{2L}{M_P^2} ~,
\end{eqnarray}
where $M_P$ and $M_*$ are the Planck and the fundamental scale mass parameters,
respectively.
For future convenience, we define $\kappa_*$ such that it has length
dimension unlike the definition in ordinary 5D SUGRA.

With the fifth dimension compactified on a $S^1/Z_2$ orbifold,
$Z_2$ parities assigned to fields in 5D N=2 theory,
consistent with SUSY transformation, are given in Table I.
\begin{table}
\begin{center}
\begin{tabular}{|c||c|c|c|c|c|c|c|c|c|c|c|c|}
\hline
$+$ &$e_m^{\;\;a}$ &$e_{55}$ &$A_5$ &$\psi_{mR}^{1}(\psi_{mL}^{2*})$
&$\psi_{5R}^{2}(\psi_{5L}^{1*})$
&$v_{a5}$ &$\zeta$ &$C$ &$V_m^3$ &$V_5^{1,2}$ &$t^{1,2}$
&$V_a$ \\
\hline
$-$  &$e_5^{\;\;a}$ &$e_{m 5}$ &$A_m$ &$\psi_{mR}^{2}(\psi_{mL}^{1*})$
&$\psi_{5R}^{1}(\psi_{5L}^{2*})$
&$V_m^{1,2}$ &$V_5^3$ &$v_{ab}$ &$t^3$
&$V_{\dot{5}}$ &$\varphi$ &$\chi$
\\
\hline
\end{tabular}
%\end{center}
%\end{table}
\vskip 0.3cm
{\bf Table I.~} $Z_2$ parity assignments for the fields in D=5, N=2
pure SUGRA.
\end{center}
\end{table}
Since only the even parity particles contain massless modes,
truncation of the heavy Kaluza-Klein (KK) modes
reduces the 5D, N=2 SUGRA to 4D, N=1 SUGRA of \cite{sohnius}.
The relations between the auxiliary fields in 4D N=1 SUGRA and
those in the ``minimal multiplet'' of 5D, N=2 SUGRA are
given by \cite{zucker3}
\begin{eqnarray} \label{relb}
b_a&=&v_{a\dot{5}} ~, \\
a_m&=& -\frac{1}{2}( V_m^3 - \frac{2}{\sqrt{3}}
\widehat{F}_{m5}e^5_{\dot{5}}+4e_m^av_{a\dot{5}})~, \\
S&=& C-\frac{1}{2}e_{\dot{5}}^5(\partial_5
t^3-\bar{\lambda}\tau^3\psi_5+V_5^1 t^2-V_5^2t^1) ~.    \label{rele}
\end{eqnarray}

The 5D N=2 hypermultiplet~\cite{zuckerth} could be extended
from the 4D N=2 hypermultiplet~\cite{dewit}.
Since there is no field carrying vector indices in the hypermultiplet,
the structures of the hypermultiplets in D=4,5,6 are almost the same.
The bosonic part of the corresponding 5D N=2 off-shell Lagrangian
is \cite{zuckerth}
\begin{eqnarray} \label{hyper}
{\cal L}_{HYP}&=&\frac{1}{2L}~d_\beta^\alpha\bigg[
~\frac{1}{2}{\cal D}_A A^j_\alpha {\cal D}^A A^\beta_j
+\frac{1}{2}F^j_\alpha F^\beta_j\bigg(1-\frac{4}{3}A_A A^A\bigg) \\
&&+A^j_\alpha A^\beta_j\bigg(\frac{1}{8}R(\hat{\omega})_{AB}{}^{AB}
+\frac{\kappa^2}{12}F_{AB}F^{AB}
-2\kappa C
-10\kappa^2\vec{t}^2-\frac{\kappa^2}{2}v_{AB}v^{AB}\bigg)
\bigg] \nonumber  ~,
\end{eqnarray}
where $j(=1,2)$ is the $SU(2)_R$ index as before, and
$\alpha(=1,2,\cdots, 2r)$ is a $USp(p,q)$ ($p+q=2r$) index, and $A$, $B$ are
5D Lorentz indices.
As in the pure SUGRA case, the linear dependence of the Lagrangian on $C$
must be resolved using the nonlinear multiplet.
Equation (\ref{constraint}) eliminates the couplings of $A^j_\alpha A^\beta_j$
with $\vec{t}^2$, $R_{AB}\,^{AB}$, and so on in the Lagrangian (\ref{hyper}).

For $r=1$, $USp(2,0)=SU(2)$,
$~d_\beta^\alpha=-\epsilon^{\alpha\gamma}\epsilon_{\gamma\beta}
=\delta_\beta^\alpha$.
The component fields in the hypermultiplet fulfill
the reality constraints \cite{zuckerth,dewit}.
For the scalar field, for instance,
\begin{eqnarray}
A^j_\alpha\equiv A^{\alpha *}_j=\epsilon^{jk}\rho_{\alpha\beta}A^\beta_k ~,
\end{eqnarray}
where $\rho_{\alpha\beta}$ is proportional to $\epsilon_{\alpha\beta}$ and
satisfies
\begin{eqnarray}
|{\rm det}\rho|^2=1 ~,~~{\rm and} ~~
\rho^{\alpha\beta}=\rho_{\alpha\beta}^* ~.
\end{eqnarray}
The auxiliary field $F^i_\alpha$ is similarly constrained.
The fermion $\xi^\alpha$ satisfies the symplectic Majorana condition,
\begin{eqnarray}
\overline{\xi_\alpha}\equiv (\xi^\alpha)^\dagger\gamma^0
=\rho_{\alpha\beta}\xi^{\beta T}C ~,
\end{eqnarray}
where $C={\rm diag}(-i\tau^2,-i\tau^2)$.

The hypermultiplet is split under $Z_2$ symmetry into two chiral multiplets
with even and odd parities, respectively.
The (bulk) chiral multiplet with
even parity can take part in the superpotential on the 4 dimensional branes.
On the other hand, the (bulk) chiral fields with odd parity vanish on the
branes and do not possess massless modes.
Consequently, they are neglected in the low energy physics.
The surviving fields at low energy are
$A^1_1$ (or $A^{2*}_2$), $\xi_{1L}$ (or $\xi_{2R}^*$)
and $F^1_1$ (or $F^{2*}_2$).

In the more general case, the Lagrangian could be written as a non-linear sigma
model, as shown in the on-shell formalisms of \cite{onshell}.
The kinetic term is then
${\cal L}_{kin}=h_{uv}(\phi)\partial_M\phi^u\partial^M\phi^{v}$, where
$h_{uv}(\phi)$ ($u,v=1,2,\cdots, 4r$) is a metric on a quaternionic manifold
with coordinates $\phi^u$.

A 4D chiral multiplet consists of scalar and fermion fields with
\begin{eqnarray} \label{chiral}
A=(A,~B;~\psi;~F,~G) ~,
\end{eqnarray}
where $A$, $B$ are dynamical real scalars, $\psi$ is a chiral fermion,
and $F$, $G$ are real auxiliary fields.
Thus bosonic degrees of freedom are the same as those of fermions.
In this paper, the first member in a chiral multiplet such as $A$
in Eq.~(\ref{chiral}) will be used, if necessary,
to designate the chiral multiplet or superfield to which it belongs.
To each chiral multiplet, a weight $w_i$ ($U(1)_R$ charge) is assigned.

According to the local tensor calculus \cite{sohnius,west},
the canonical kinetic part of the off-shell Lagrangian for the chiral multiplet
is given by the ``D-density'' of the product, $-[A_i\times A_i]_D/4$,
and its bosonic part is \cite{zucker3,zuckerth,sohnius}
\begin{eqnarray} \label{realkinetic}
{\cal L}_{4dCHI}
&=&\sum_i\bigg[~\frac{1}{2}\bigg(
\hat{{\cal D}}_mA_i\hat{{\cal D}}^mA_i+\hat{{\cal D}}_m B_i\hat{{\cal D}}^mB_i
+F_i^2+G_i^2\bigg)   \\
&&-4\kappa t^2(F_iA_i-G_iB_i)
-4\kappa t^1(F_iB_i+G_iA_i)
+2\kappa b^m(A_i\hat{{\cal D}}_mB_i-B_i\hat{{\cal D}}_mA_i) \nonumber \\
&&-\frac{\kappa}{2}(A_i^2+B_i^2)
\bigg(12w_iS-\frac{w_i}{4\kappa}R_{ab}{}^{ab}
+48\kappa (w_i-1)
[(t^1)^2+(t^2)^2]\nonumber  \\
&&-6\kappa w_ib_mb^m-8S\bigg)\bigg] ~, \nonumber
\end{eqnarray}
where $i$ labels chiral multiplets, and
$\hat{{\cal D}}_m$ is the supercovariant derivative, defined as \cite{zuckerth}
\begin{eqnarray}
\hat{{\cal D}}_m A &\equiv& \partial_m A-\bar{\psi}_m\tau^2\psi-wBa_m ~, \\
\hat{{\cal D}}_m B &\equiv& \partial_m B-\bar{\psi}_m\tau^2\gamma^{\dot{5}}\psi
+wAa_m ~.
\end{eqnarray}
We note that in the Lagrangian (\ref{realkinetic}) several
weight ($w_i$) dependent terms appear.
%
%, and
%many auxiliary fields involved in the gravity multiplet
%as well as the chiral multiplet's auxiliary fields contribute
%the 4D Lagrangian of the chiral multiplet.
%
This can be generalized to non-trivial cases,
$[f(A)\times A]_D$, where $f$ is an arbitrary function of $A$ \cite{west}.
The Lagrangian in Eq.~(\ref{realkinetic}) can be written in terms of
5D SUGRA auxiliary fields using Eqs.~(\ref{relb})--(\ref{rele}),
\begin{eqnarray} \label{compkinetic}
{\cal L}_{5dCHI}&=&\sum_{I,i}\delta(y-y_I)\bigg[
|\partial \phi_i^{I}|^2+|{\cal F}_i^{I}|^2
-4\kappa_*\phi^{I*}_i{\cal F}^{I}_i{\cal M}
-4\kappa_*\phi^{I}_i{\cal F}^{I*}_i{\cal M}^* \\ \nonumber
&&-i\frac{\kappa_*}{2}\bigg(\phi^{I}_i\partial^m\phi_i^{I*}
-\phi^{I*}_i\partial^m\phi^{I}_i\bigg)
\bigg(w_iV_m^3-\frac{2w_i}{\sqrt{3}}F_{m\dot{5}}+4(w_i-1)v_{m\dot{5}}\bigg)
\\ \nonumber
&&+\kappa_*|\phi^{I}_i|^2\bigg((3w_i-2)(2\partial_{\dot{5}}t^3
+\kappa_* {\cal M}{\cal V^*}
+\kappa_* {\cal M}^*{\cal V})-48\kappa_* (w_i-1)|{\cal M}|^2 \\ \nonumber
&&-12w_iC+8C+\frac{w_i}{4\kappa_*}R_{ab}{}^{ab}
-6\kappa_* w_iv_{a\dot{5}}v^{a\dot{5}}  \\ \nonumber
&&+\frac{\kappa_*}{4}w_i^2(V_m^3-\frac{2}{\sqrt{3}}F_{m\dot{5}}+4v_{m\dot{5}})
(V^{m3}+\frac{2}{\sqrt{3}}F^{m\dot{5}}-4v^{m\dot{5}}) \\ \nonumber
&&+2\kappa_* w_iv^{m\dot{5}}(w_iV_m^3-\frac{2}{\sqrt{3}}F_{m\dot{5}}
+4v_{m\dot{5}}) \bigg) \bigg] ~,
\end{eqnarray}
where we complexifed some bosonic degrees for future convenience,
\begin{eqnarray}\label{complexb}
\phi&\equiv&\frac{1}{\sqrt{2}}(A+iB) ~,\\
{\cal F}&\equiv&\frac{1}{\sqrt{2}}(F-iG) ~,\\
{\cal M}&\equiv& t^2+it^1 ~,\\
{\cal V}&\equiv& V_{\dot{5}}^1-iV_{\dot{5}}^2 \label{complexe} ~.
\end{eqnarray}
Note that in Eq.~(\ref{compkinetic}),
the expansion parameter is not ``$\kappa$'' but ``$\kappa_*$''
defined in Eq.~(\ref{fundamental}).
Since the Lagrangian (\ref{compkinetic}) describes the dynamics of scalar
fields localized on the brane, a delta function appears as an overall factor.
$I=1,2$ indicate the two brane locations at $y=0$ and $y=L$, respectively.

Next we briefly review how the superpotential is constructed in the 4D SUGRA,
which would enable us to derive the relevant scalar potentials.
In the local tensor calculus \cite{sohnius,west},
a product ``$\cdot$'' is defined such that
the product, $A_3=A_1\cdot A_2$ of two chiral multiplets
of weights $w_1$ and $w_2$, respectively, yields
another chiral multiplet with weight $w_3=w_1+w_2$,
and components
\begin{eqnarray} \label{ruleb}
A_3&=&A_1A_2-B_1B_2 ~,\\
B_3&=&A_1B_2+A_2B_1 ~, \\
\psi_3&=&(A_1-\gamma_5 B_1)\psi_2+(A_2-\gamma_5 B_2)\psi_1 ~, \\
F_3&=&A_1F_2+B_1G_2+F_1A_2+G_1B_2+\bar{\psi}_1\psi_2 ~, \\
G_3&=&A_1G_2-B_1F_2+G_1A_2-F_1B_2-\bar{\psi}_1\gamma_5\psi_2 ~. \label{rulee}
\end{eqnarray}
These product rules are valid in the locally supersymmetric case as well as
in global supersymmetry.
Using the rules, we can define superpotentials such as $A_1\cdot A_2$,
$A_1\cdot A_2\cdot A_3,~\cdots$.

By examining the SUSY transformations of members in a chiral multiplet,
one can find a supersymmetric invariant, ``F-density,''
which is useful for constructing the Lagrangian.
A supersymmetric invariant up to total derivative terms
in SUGRA is given by \cite{zucker3,sohnius,west}
\begin{eqnarray}\label{fdensity}
[A]_F =
F+i\kappa\bar{\psi}^2_m\gamma^m\psi
+\frac{\kappa^2}{2}\bar{\psi}_m\tau^2 \gamma^{mn}(A+\gamma^{\dot{5}}B)\psi_n
-12\kappa t^2A-12\kappa t^1B ~,
\end{eqnarray}
with weight equal to 2~\cite{sohnius}.
The Pauli matrix $\tau^2$ contracts the $SU(2)_R$ indices of the gravitino.
The first term in Eq.~(\ref{fdensity}) is present also
in the globally supersymmetric case, while
the other terms are the SUGRA corrections.
In particular, if $A$ and/or $B$ acquire vacuum expectation values (VEVs),
the third term could yield the gravitino mass term by absorbing
the chiral fermion's degree of freedom of the second term.
Thus, for a given superpotential $W(\phi_1,\phi_2,\cdots)$,
the Yukawa interactions and scalar potentials are
read off using Eqs.~(\ref{ruleb})--(\ref{rulee}) and (\ref{fdensity}),
and it can be checked that the bosonic part is given by
\begin{eqnarray}
{\cal L}_{4dYUK}=\bigg[\sum_zW_z{\cal F}_z
-12\kappa W{\cal M}
+{\rm h.c.}\bigg] ~,
\end{eqnarray}
where every bosonic field is complexified
using Eqs.~(\ref{complexb})--(\ref{complexe}), ${\cal F}_z$ is the
auxiliary field involved in $\phi_z$, and
\begin{eqnarray}
W_z\equiv \frac{\partial W}{\partial \phi_z} ~.
\end{eqnarray}

\section{Effective 4D scalar potential}

To illustrate the emergence of non-universal soft masses,
let us consider a model with
a visible ($V$) and a hidden ($H$) hypermultiplet in the bulk,
and two visible and some hidden chiral multiplets localized at each brane.
The chiral multiplet from the bulk hypermultiplets with even parity
$\Phi^V$ comprises a trilinear superpotential
together with two visible chiral multiplets $\phi^{1V}_1$ and $\phi^{1V}_2$,
localized at B1, while
the hidden sector multiplets $\Phi^H$ and $\phi^{1H}_x$ ($x=a,b,\cdots$)
make up another superpotential at B1,
\begin{eqnarray}
W^{1V}&=&Y^{1V}_{12}~\Phi^V\phi_1^{1V}\phi_2^{1V},~~~~~~~~~~
W^{1H}=W^{1H}(\Phi^H,\phi^{1H}_a,\phi^{1H}_b,\cdots) ~.
\end{eqnarray}
Similarly, the two (bulk) chiral multiplets,
the two localized visible multiplets, $\phi^{2V}_1, \phi^{2V}_2$, and
hidden chiral multiplets $\phi^{2H}_a$ ($x=a,b,\cdots$) constitute
the following superpotentials at B2,
\begin{eqnarray}
W^{2V}&=&Y^{2V}_{12}~\Phi^V\phi_1^{2V}\phi_2^{2V},~~~~~~~~~~
W^{2H}=W^{2H}(\Phi^H,\phi^{2H}_a,\phi^{2H}_b,\cdots) ~.
\end{eqnarray}
We assign weights 2, 0, and 2/3 to $\Phi^V$, $\phi^{IV}_{i}$ ($I,i=1,2$),
and $\phi^{IH}_{x}$, respectively, for simplicity of the model.
Other weight assignments that do not change our essential conclusions
are also possible.

The bosonic part of the locally supersymmetric off-shell Lagrangian
for the Yukawa interactions is
\begin{eqnarray}\label{ykw}
{\cal L}_{YUK}&=&\sum_{I,S,z}\delta (y-y_I)\bigg[W^{IS}_z{\cal F}^{IS}_z
-12\kappa_* W^{IS}{\cal M}
+{\rm h.c.}\bigg] ~,
\end{eqnarray}
where $I$ ($=1,2$) denotes the brane locations ($y_1=0, y_2=L$),
$S$ stands for visible or hidden sectors ($S=V,H$),
$z=\Phi^V,1,2$ for $S=V$, and $z=\Phi^H,a,b,\cdots$ for $S=H$.
To get a scalar potential for the dynamical fields,
we replace the auxiliary fields in the Lagrangians
(\ref{sugra}), (\ref{hyper}), (\ref{compkinetic}), and (\ref{ykw}),
using the equations of motion.
The equations of motion for ${\cal {F}}$ fields give
\begin{eqnarray} \label{fbrane}
{\cal F}^{IS}_i&=&-(W_i^{IS})^*+4\kappa_*\phi^{IS}_i{\cal M}^* ~, \\
{\cal F}_\Phi^{IS}&=&-\sum_{I}2\Delta(y_I)(W^{IS}_\Phi)^*
\bigg(1-\frac{4}{3}\kappa^2A_5A^5\bigg)^{-1} ~, \label{fbulk}
\end{eqnarray}
where $i$ is $1, 2$ for $S=V$, and $a,b,\cdots$ for $S=H$, and
\begin{eqnarray}
\Delta (y_I)\equiv L\times \delta(y-y_I)~.
\end{eqnarray}
In Eq.~(\ref{fbulk}), we see that the bulk field ${\cal F}^S_\Phi$
gets contributions from both localized brane ``sources.''
Let us insert the above expressions into the orginal Lagrangian
and also eliminate the auxiliary field ${\cal M}$, whose
equation of motion gives
\begin{eqnarray} \label{m}
{\cal M}&=&-\sum_{I=1,2}\frac{2\kappa_*}{3}\Delta(y_I)
\bigg[1+\frac{4\kappa_*^2}{3}\Delta(y_I)\sum_{i}|\phi^{IV}_i|^2
%+\frac{1}{2}\sum_x|\phi^{IH}_x|^2(2-3w_x)\bigg)
\bigg]^{-1} \\
&&~~~~~\times\bigg[
\bigg(W^{IV}+W^{IH}\bigg)+\frac{\kappa_*}{2}\sum_{i}|\phi^{IV}_i|^2
%+\frac{1}{2}\sum_{x}|\phi^{IH}_x|^2(2-3w_x)\bigg)
{\cal V}
\bigg] ~. \nonumber
%&&~~~~~\times\bigg[1+\frac{4\kappa_*^2}{3}\Delta(y_I)\sum_{i}|\phi^{IV}_i|^2
%+\frac{1}{2}\sum_x|\phi^{IH}_x|^2(2-3w_x)\bigg)
%\bigg]^{-1} ~.
\end{eqnarray}

In Eq.~(\ref{m}), the delta function $\Delta (y_I)$
is present in the denominator as well as the numerator,
which seems surprising at first.
However, to couple the 5D gravity multiplet
supersymmetrically to 4D chiral fields,
the delta function couplings are inevitable.
In this paper, we will take a pragmatic approach to the delta function;
we regard the $\delta (y-y_I)$ functions as {\it finite} walls
of height $M_*$ with a small but non-vanishing ``thickness'' $1/M_*$.
(The exact shape of the delta function would be determined by a fundamental
theory, but it would not affect the low energy physics.)
Despite their thickness, both walls are assumed to be sufficiently far
apart from each other, so that
\begin{eqnarray}\label{deltamul}
\Delta (y_I)\times \Delta (y_J) = 0 ~ ~~~~~~{\rm for}~~~I\neq J~.
\end{eqnarray}
Due to Eq.~(\ref{deltamul}), ${\cal M}$ in Eq.~(\ref{m})
is given by the sum of the contributions from the two branes.
In a more fundamental theory, the ``blow-up'' of the delta function could
be cured and excitations of localized fields on the branes
in the $y$ direction would be possible.
The apparently divergent operator in the denominator in Eq.~(\ref{m})
is now suppressed,
\begin{equation}
\kappa_*^2\Delta (y_I)~|\phi^{IV}_i|^2~\sim ~\frac{L}{M_*}~|\phi^{IV}_i|^2 ~,
\end{equation}
and we can treat it perturbatively
unless $\langle\phi^{IV}_i \rangle \sim M_*$,
as in ordinary 4D supergravity.
In our case, we take
\begin{eqnarray}
L~\sim~10^{-16}~{\rm GeV}^{-1} ~.
\end{eqnarray}

For the hidden sector fields and superpotentials
we assume the VEVs \cite{nilles}
\begin{eqnarray} \label{bvev}
%&&\langle~\phi_x^{IH}~\rangle=a_x^IM_*~, \\
%&&\langle~\Phi^{IH}~\rangle=b^IM_*~, \\
&&\langle~W^{IH}~\rangle=-m_IM_*^2 ~, \label{hiddenvev}\\
&&\langle~W^{IH}_z ~\rangle
=(b^{I}_z)^*m_IM_* ~. \label{evev}
\end{eqnarray}
For convenience, we define the following expressions:
\begin{eqnarray}
&&W^{IV}~\equiv~W^I ~, ~~~~~W_{\Phi}^{IV}\equiv W_{\Phi}^{I} ~,
~~~~~\sum_i|W^{IV}_i|^2\equiv |W^I_i|^2 ~,  \\
&&\sum_i|\phi_i^{IV}|^2\equiv |\phi_i^I|^2 ~,
~~~~~\sum_{z}|b_z^I|^2\equiv |b_z^I|^2 ~.
\end{eqnarray}
%
%%&&\sum_x|a_x^I|(2-3w_x)\equiv |a_x^I|^2(2-3w_x) ~, ~~~~~
%%\sum_{x}|b_x^I|^2\equiv |b_x^I|^2 ~,  \\
%%&&Q^I\equiv 1+\frac{2}{3}\Delta(y_I)|a_x^I|^2(2-3w_x) ~,
%\end{eqnarray}
%Then, Eq.~(\ref{m}) can be approximately given by
%\begin{eqnarray} \label{approxm}
%{\cal M}&=&-\sum_{I=1,2}\frac{2\kappa_*\Delta(y_I)}{3Q^I}
%\bigg[\bigg(W^I+m_IM_*^2\bigg)
%+\frac{\kappa_*}{2}\bigg(|\phi^{I}_i|^2
%+\frac{M_*^2}{2}|a_x^I|^2(2-3w_x)\bigg){\cal V} \nonumber \\
%&&~~~~~-\frac{4\kappa_*^2}{3Q^I}\Delta(y_I)\bigg(W^{I}+m_IM_*^2\bigg)|\phi^{I}_i|^2
%\\
%&&~~~~~-\frac{2\kappa_*^3}{3Q^I}\Delta(y_I)\bigg(|\phi^{I}_i|^2
%+\frac{M_*^2}{2}|a_x^I|^2(2-3w_x)\bigg)|\phi_i^I|^2{\cal V}
%\bigg]  ~+~O(\kappa_*^6)~+~\cdots \nonumber ~.
%\end{eqnarray}
%In Eq.~(\ref{approxm}), we keep only the terms of order upto $O(\kappa^4)$.
%It is, however, enough to examine the soft scalar masses and
%A terms.
%The $O(1)$ contributions to the denomenator in Eq.~(\ref{m}) by
%the hidden sector fields' VEV of order $M_P$ are absorbed
%in the factor `$Q^I$'.
%
After inserting Eq.~(\ref{m}) and Eqs.~(\ref{bvev}) and (\ref{evev})
into the original Lagrangian, assembling the results from the auxiliary
fields ${\cal F}_i^{IS}$ and ${\cal F}_{\Phi}$
in Eqs.~(\ref{fbrane}) and (\ref{fbulk}),
we obtain the localized F term scalar potentials for $\Phi$ and $\phi^I_i$
in 5D supergravity,
\begin{eqnarray} \label{5dpot}
V_{5d}(\Phi,\phi^I_i)&=&\sum_{I=1,2}\frac{\Delta(y_I)}{L}\bigg[|W^I_i|^2
+2\Delta(y_I)|W_\Phi^I|^2+|b_z^I|^2(m_IM_*)^2
%\bigg(
%1+\frac{4\kappa_*^2}{3}A_5A^5-\frac{16\kappa_*^4}{9}(A_5A^5)^2\bigg)
\\
&&-\frac{8\kappa_*^2}{3}\Delta(y_I)|W^I-m_IM_*^2|^2
\bigg(1-\frac{4\kappa_*^2}{3}\Delta(y_I)|\phi_i^I|^2\bigg)
\bigg]  +O(\kappa_*^6) \nonumber ~,
\end{eqnarray}
where we retain only terms up to $O(\kappa^4)$ for
examining the soft scalar masses and $A$ terms.
Elimination of ${\cal V}$ gives rise to corrections
of order $O(\kappa^6)$ to the potential (\ref{5dpot}),
and from the equations of motion for $V_m^3$ and $v_{m\dot{5}}$, the
derivative interaction terms between $\phi_i^I$s and $F_{m\dot{5}}$ are
derived.
We should note here that in the localized potential,
there is no cross term between the two brane contributions
due to the delta function property in Eq.~(\ref{deltamul}).
Thus the potential at B1 is associated only with bulk scalar $\Phi$ and
B1 brane fields $\phi^1_1$ and $\phi^1_2$, while the potential at B2
contains $\Phi$ and B2 brane fields $\phi^2_1$ and $\phi^2_2$.

We note the appearance of $\Delta(y_I)$ dependent couplings in the potential
(\ref{5dpot}).   In momentum space,
\begin{eqnarray}\label{deltaexp}
\delta(y-y_I)=\frac{1}{2L}\sum_{n=-\infty}^\infty e^{i(n\pi/L)(y-y_I)}
=\frac{1}{2L}\sum_{k_5=-\infty}^\infty e^{ik_5(y-y_I)} ~,
\end{eqnarray}
which is used in the Feynman rule for the delta function coupling for
calculating certain processes \cite{peskin,gherg}.
%The expression in momentum space
%plays the role of a Feynman rule for the coupling
%when we calculate some processes.
To obtain an effective 4D theory,
we truncate the KK modes and introduce the cuttoff,
$\Lambda_{c}\lapproxeq 1/L$ ($\sim 10^{16}$ GeV).
Thus the summation in Eq.~(\ref{deltaexp}) should be implemented
only for $k_5=0$ or $n=0$, so that
$\Delta(y-y_I)$ in the potential (\ref{5dpot})
reduces to $1/2$ in low energy physics,
{\it independent of regularization schemes for the delta function}.
Cross interaction terms
between fields localized on two different branes
could be generated at loop level below $\Lambda_c\sim 1/L$, while
above $\Lambda_{c}\sim 1/L$,
the effective theory is 5D SUGRA and the
``finite'' delta function (or the finite wall)
properties like Eq.~(\ref{deltamul}) are restored.

Before deriving the effective 4D potential, let us discuss
the delta function coupling for a moment.
In Ref.~\cite{peskin},
the supersymmetric coupling between 5D bulk super Yang-Mills and
charged chiral matter localized on 4 dimensional brane was studied.
In this model too, a delta function squared coupling
appears after eliminating the auxiliary field $X^{3a}$.
The 5D gauge multiplet contains a dynamical scalar $\Phi^a$
with odd parity under $Z_2$ symmetry, so that
$X^{3a}-\partial_5\Phi^{a}$ should be identified,
from their SUSY transformation properties,
with the auxiliary field $D^a$ appearing in the 4D gauge multiplet.
The delta function coupling is interpreted as a
counterterm to divergences arising from
the derivative interactions $\partial_5\Phi^a$ of an odd parity
{\it scalar} with other chiral fields.
However, such a dynamical odd parity scalar field is not found in the
gravity multiplet as seen from Table I.
Moreover, even if present,
the scalar with odd parity would be decoupled at low energy.\footnote{In
Ref.~\cite{gherg}, a 5D SUGRA coupled to a 4D super Yang-Mills
was considered,
where a delta function squared coupling appears also in the 5D Lagrangian.
Since a delta function squared coupling in this case has a vector-like
interaction involving gauginos, the odd parity
vector field (graviphoton) plays the role of $\Phi^a$
of Ref.~\cite{peskin}.  Here we intend to derive the {\it scalar} potential,
so the vector field is not relevant. }
In Ref.~\cite{falkowski}, an additional odd parity scalar included in
a hypermultiplet plays the role of $\Phi^a$ in Ref.~\cite{peskin}.
But SUSY transformations of the 5D SUGRA auxiliary fields
do not necessarily require additional fields
for proper identification with the 4D auxiliary fields.
Most pressing of all, in our model,
such a method appears to be unable to control every delta function
coupling.\footnote{
This results from the fact that in general
the Lagrangian of the 4D chiral multiplet
contains quadratic terms in ${\cal M}$, while in previous examples
the associated auxiliary fields, namely $D^a$ in Ref.~\cite{peskin} and
$b_m$ in Ref.~\cite{gherg}, appear in
only linear form in the corresponding brane Lagrangian.  }
Thus we have adopted a different approach to the delta function coupling
in which it is regarded as a wall with height $M_*$
and width $1/M_*$.
Then, even if the wall of the delta function is
as high as the fundamental scale $M_*$ or even the Planck scale,
all delta function couplings are in turn
small since they are always accompanied by the SUGRA parameter $\kappa_*$.
However, in the low energy effective theory,
the delta function couplings are not problematic in any case
if the introduced cutoff ($1/L\sim 10^{16}$ GeV) is high enough to
suppress any induced non-renormalizable operators.

The effective 4D potential is obtained
by integrating $V_{5d}$ over
the fifth direction $-L\leq y\leq L$,
\begin{eqnarray}\label{4dpot}
V_{4d}&=&\sum_{I=1,2}(m_IM_P)^2\bigg(|b_z^I|^2
-\frac{4}{3}\bigg)
+|W^I_i|^2+|W_{\Phi}^I|^2 \nonumber \\
&&~~~~~+\frac{4m_I}{3}(W^I+W^{I*})
+\frac{8(m_I)^2}{9}|\phi_i^I|^2
 \\
&&~~~~~
+{\rm non-renormalizable}~{\rm interactions} ~.   \nonumber
\end{eqnarray}
In Eq.~(\ref{4dpot})
the cosmological constant term and the ordinary F term potential of
globally supersymmetric theory appears in the first line, the SUSY breaking
$A$ terms and soft scalar mass terms are in the second line.
The non-renormalizable terms are suppressed by powers of $1/M_*$.
As in ordinary 4D supergravity,
a vanishing cosmological constant requires fine-tuning.
Thus we get a theory with softly broken SUSY
but a vanishing cosmological constant
through the fine-tuning,
which is impossible in globally supersymmetric theory.
In contrast to the 4D case,
the global SUSY F term potential is divided into two parts that
depend on the locations of the associated fields \cite{kyae}.
Note that the potential in Eq.~(\ref{4dpot}) does not allow
a non-zero VEV for scalar fields, so the internal symmetry cannot be
spontaneously broken at the fundamental scale.
With $\sum_{i}W_i^I\phi_i^I=2W^I$,
this can be proved by showing that while the minimum value of the potential
always becomes non-zero and positive
when a non-zero VEV for any scalar is assumed,
it is vanishing for zero VEVs of all of the scalar fields \cite{nilles}.
It is possible that the internal symmetry breaks down radiatively
at low energy, as in ordinary 4D SUGRA models.

From the scalar curvature terms in Eqs.~(\ref{sugra})
and (\ref{compkinetic}), we see that
the effective 4D potential in Eq.~(\ref{4dpot}) is not yet
in the Einstein frame.
However, the canonical normalization of the effective 4D gravity
just rescales the Yukawa coeffecients and soft masses
in Eq.~(\ref{4dpot}) by a small amount,
and adds non-renormalizable interactions.

As seen in Eq.~(\ref{4dpot}), a soft mass term of the bulk scalar is not
generated at tree level under the formalism employed in this paper.
It is basically because the terms $\kappa^2 A^j_\alpha A^\beta_j |\vec{t}|^2$
(or $\kappa^2|\Phi|^2|{\cal M}|^2$) and $A^j_\alpha A^\beta_j R_{AB}\,^{AB}$
(or $|\Phi|^2R_{AB}\,^{AB}$) are cancelled out
from the Lagrangian (\ref{hyper}).
If the term $\kappa^2|\Phi|^2|{\cal M}|^2$ exists in the Lagrangian,
the bulk scalar could couple to the superpotentials $W^I$
on the branes by Eq.~(\ref{ykw})
when eliminating the auxiliary field ${\cal M}$.
However, a more general formalism
might allow the soft mass term of the bulk scalar,
$\kappa^2|\Phi|^2|\kappa_*\langle W^{IH}\rangle|^2$
($\sim|\Phi|^2m_{I}^2M_*^2/M_P^2$)
at tree level, since it apparently does not break any symmetry.
Moreover, when the term $|\Phi|^2R_{AB}\,^{AB}$ is present
(generally when the metric on a quaternionic manifold with coordiates
$A^j_\alpha$ has a non-trivial form),
a soft mass term of the bulk scalar can be also generated through the coupling
between the bulk scalar and the F term of a hidden sector superfield
like $\kappa^2|\Phi|^2|\langle {\cal F}_z^{IH}\rangle|^2$
($\sim |\Phi|^2 m_{I}^2M_*^2/M_P^2$) in the Lagrangian.
Such a coupling could be obtained after canonically normalizing
the 5D gravity kinetic term.
The coupling could be important in phenomenology~\cite{bs}.
Here we will just follow the formalism of Ref.~\cite{zucker1}.
In this formalism, a soft mass term of the bulk scalar can be generated
at one loop.   We will discuss it later.

\section{Non-universal SUSY breaking soft parameters}

In the potential (\ref{4dpot}), we note that
the soft mass term of the bulk scalar $\Phi$ is not generated by
localized SUSY breaking at tree level.
On the other hand, fields localized on the branes $\phi^I_i$ obtain
their soft masses only from their associated branes.
Since the mass parameters $m_I$ are different in general,
\begin{eqnarray}
m_I~\neq ~m_J ~~~~~~~{\rm for}~~I~\neq ~J ~,
\end{eqnarray}
$A$ terms and soft scalar masses of brane fields
are also different at tree level,
unless the relevant brane scalar fields live on the same brane.
This consequence is definitely different from that of the ordinary 4D
SUGRA scenario.
The latter corresponds
to taking the limit $L\rightarrow 0$ from the start and neglecting the
delta functions in our model.
Then, the auxiliary field ${\cal M}$ does not discern the location
of fields, and cross terms between the two brane contributions appear,
for example, from $|{\cal M}|^2$ as well as from ${\cal M}$.
Thus all of the $A$ terms and soft scalar masses receive contributions from
both $\langle W^{1H}\rangle$ and $\langle W^{2H}\rangle$,
and universality of SUSY breaking soft parameters is observed.

The different SUSY breaking masses are confirmed
by cross-checking the gravitino mass terms.
With Eqs.~(\ref{ruleb})--(\ref{rulee}),
the third term in Eq.~(\ref{fdensity}) provides the localized gravitino masses,
\begin{eqnarray}
m_{3/2,I}~\sim~\kappa_*^2 \langle ~W^{IH}~\rangle ~\sim ~m_I ~.
\end{eqnarray}
Therefore the gravitino acquires mass $m_1$ from B1 and $m_2$ from B2,
%even if heavier mass is taken in the effective 4D theory.
so that the SUSY breaking scales are
$O(m_1)$ and $O(m_2)$ at B1 and B2, respectively.

Although the SUSY breaking effects are generated through the direct coupling
appearing in SUGRA at tree level and they are localized on the corresponding
branes, they can be transmitted through ``loops''
to the bulk and even to the other branes
below the compactification scale \cite{gherg}.
The off-shell Lagrangian of the hypermultiplet contains
the interaction term with
the gravitino and $t^2$ \cite{zuckerth},
\begin{eqnarray}\label{loop1}
{\cal L}_{HYP}&\supset&-\frac{\kappa^2}{L}A^\alpha\tau^2~
\overline{\zeta}_\alpha\gamma^m\psi_m~t^2  ~, \label{loop2}
%{\cal L}_{5dCHI}&\supset&-\delta(y-y_I)~4i\kappa_*^2\sqrt{2}~
%\phi_i^I~\overline{\psi_i^I}\gamma^m\psi_m ~t^2 ~,
\end{eqnarray}
which survives under the $S^1/Z_2$ compactification.
The VEV $\langle t^2\rangle=\langle {\cal M}\rangle\sim
\Sigma_I\Delta(y_I)m_IM_*$ from SUSY breaking at the branes
makes a bulk scalar's soft mass generated at one loop, so that \cite{gherg}
\begin{eqnarray}\label{loopmass}
\delta m_{\Phi}^2&\sim&\sum_I\frac{1}{16\pi^2}\bigg(\frac{m_IM_*}{M_P^2}\bigg)^2
\frac{1}{L^2} ~,
%\delta m_{\phi^I_i}^2&\sim&\sum_{J}\frac{1}{16\pi^2}\bigg(\frac{m_J}{M_*}\bigg)^2
%\frac{1}{L^2} ~,
\end{eqnarray}
where $1/L$ is the cutoff $\Lambda_c$.
Note that for $1/L\sim 10^{16}$ GeV, $M_*\sim 10^{17}$ GeV,
and $m_1\sim 10$ TeV,
$\delta m_{\Phi}$ is of order a few GeV, which is quite small compared to
other SUSY breaking soft masses.

We could exploit these phenomena for resolving the notorious flavor changing
problem in SUGRA.
We assume that

\noindent $\bullet$ the first two MSSM generations reside at B1;

\noindent $\bullet$ the third generation is at B2;

\noindent $\bullet$ the two Higgs multiplets as well as gravity and
gauge multiplets are in the bulk;

\noindent $\bullet$ SUSY breakings arise from two hidden sectors
localized on branes, and are subsequently transmitted by gravity.
The SUSY breaking scale at B1 should be suitably higher than
the breaking scale at B2.

Let us now consider gaugino masses.
In Ref.~\cite{zucker2},
the gaugino can become massive when SUSY is broken in the hidden sector,
\begin{eqnarray} \label{gauginomass}
{\cal L}_{SYM}^{\rm bulk}=\frac{1}{2L}{\rm Tr}\bigg[\frac{-1}{4}F_{MN}F^{MN}
%+i\bar{\lambda}\gamma^m{\cal D}_m\lambda
+\cdots
-2\kappa ~\bar{\lambda}\vec{\tau}\lambda ~\vec{t} + \cdots \bigg] ~,
\end{eqnarray}
where $\lambda$ is the gaugino which satisfies
the symplectic Majorana condition.
The coupling of the auxiliary field ``$\vec{t}$''
to $\bar{\lambda}\vec{\tau}\lambda$ means that
the visible sector gaugino becomes massive either
from the  hidden sector gaugino condensates
$\langle \bar{\lambda}^H\vec{\tau}\lambda^H\rangle\neq 0$,
or if the hidden sector
superpotential on brane obtains a VEV $\langle W^H\rangle\neq 0$.
After orbifolding, only $t^1$ and $t^2$ (or $\tau^1$ and $\tau^2$)
contributions survive at low energy, and yield the 4D Majorana mass term.
With $\langle t^2\rangle=\langle {\cal M}\rangle\sim \sum_I\Delta(y_I)m_IM_*$
from Eqs.~(\ref{m}) and (\ref{hiddenvev}),
the gaugino mass at low energy is given
\begin{eqnarray}
m_{1/2}~\sim ~m_1\frac{M_*}{M_P} ~.
\end{eqnarray}
Note that the gaugino mass is lighter for a larger extra dimension.
With $m_1=10$ TeV and $M_*=10^{17}$ GeV (or $1/L\sim 10^{16}$ GeV),
the gaugino mass is of order 1 TeV, which manages to preserve
the gauge hierarchy solution.
This could be decreased to a few hundred GeV for $1/L\sim 10^{12}$ GeV.

Let us consider another possibility for lowering the gaugino mass.
We can introduce an additional set of the localized gauge multiplet
($A_m$, $\lambda$, $D$) at B2, which respects the same gauge symmetry
as the bulk gauge multiplet does.
%
%Additional localized gauge multiplet
%does not spoil either gauge and supersymmetry.
Consider the Lagrangian
\begin{eqnarray}
{\cal L}_{SYM}=\bigg[{\cal L}_{SYM}^{\rm bulk}+N\delta(y-L)\bigg(
-\frac{1}{4}F_{mn}F^{mn}+i\bar{\lambda}\gamma^m{\cal D}_m\lambda
+\cdots \bigg) \bigg]  ~,
\end{eqnarray}
%where a large (small) coefficient $N$ corresponds
%to a small (large) gauge coupling
%of the additional localized kinetic term.
with the same gauge coupling.
Below the compactification scale $1/L$,
the coefficient of the gauge kinetic term is given by $-(1+N)/4$.
Thus canonical normalization of the kinetic term results in
the suppressed (bulk) gauge coupling $g/\sqrt{1+N}$,
as well as a suppressed gaugino mass $m_{1/2}/(1+N)$.
Hence with $N\sim O(10)$ (and strong coupling $g\sim O(1)$)
and $1/L\sim ~10^{16}$ GeV, a gaugino mass in the 100 GeV range can
be achieved.  In such a way, the masses of the various gauginos
in MSSM also could be made non-universal.

To realize electroweak symmetry breaking in the MSSM,
both the $\mu$ term and the ``$B$'' term ($\sim \mu m_{3/2}$)
of the right magnitudes must be present.
Hence it is desirable that the $\mu$ term is located at B2.
From the weight assignments for hypermultiplets,
the ordinary bilinear $\mu$ term is not allowed in our example, so
we should assume that the $\mu$ term derives either from some trilinear
or from a
non-renormalizable interaction between two Higgs and some singlet field
localized at B2 which develops a suitable non-zero VEV.
We will not pursue this further in this paper.

%Since the two Higgs multiplets are regarded as hypermultiplets
%living in the bulk in our model,
%their soft masses, which are radiatively generated, are too small as seen above.
%Thus, the large top quark Yukawa coupling would draw the value of $m_{H_2}^2$
%upto $-(1~{\rm TeV})^2$ at $m_Z$ scale (The small initial value of $m_{H_2}^2$
%at $M_*$ just shifts below the graph of its solution to the RG equation
%with a larger initial value).
%Hence, in order to break electroweak symmetry at 100 GeV scale and
%get the $m_Z^2$ consistent with the observed value,
%the fine-tuning of order $(1~ {\rm TeV})^2$ or so
%between $m_{H_2}^2$ and $|\mu|^2$ is necessary.

The electroweak Higgs doublets reside in the bulk, and
their soft mass squareds at $M_*$ arise predominantly through
loops involving quarks and squarks
rather than loops involving fields in the gravity multiplet
as in Eq. (\ref{loopmass}).
But their values at $M_*$ are also much smaller than the weak scale.
%estimated as being of order (10 ${\rm GeV})^2$ at $M_*$.
The large top Yukawa coupling and the heavy first two generations'
soft mass squareds will drive this to
$-({\rm a~few~TeV})^2$ at $M_Z$ as in the 4D ESUSY.
%(The KK modes for large extra dimension would make the two loop effects
%by the heqvy first two generations' soft mass squareds smaller.)
Consequently, an adjustment of TeV scale parameters may be needed to
achieve the desired 100 GeV electroweak symmetry breaking scale.

Finally, mixing of the first two generations with
the third generation can arise effectively through one loop
interactions below the compactification scale
by introducing suitable heavy bulk fields belonging to
vector-like representations of the SM gauge symmetry
that couple to the ordinary MSSM particles on the branes.
For neutrino mixings, we could introduce right-handed neutrinos in the bulk.
We will not go into detail here.

\section{Conclusion}

In conclusion, through the use of off-shell SUGRA formalism
and brane world framework, we have shown that gravity mediated
SUSY breaking effects can be non-universal.
With SUSY breakings in two hidden sectors localized on distinct branes,
the gravity mediated soft masses of the bulk visible sector, gauginos, and
the gravitino result from both branes, while
SUSY breaking tree level effects in the localized visible sector fields
arise only from the hidden sector living at the same brane.
The soft masses for the bulk scalars are generated radiatively.
This enables us to overcome the SUSY flavor problem geometrically
by generating soft scalar masses $\sim$10--20 TeV
for the first two generations,
while the third generation masses are of order a TeV.
Radiative electroweak breaking can be realized in this approach, with TeV scale
quantities conspiring to realize the 100 GeV electroweak scale.

\vskip 0.3cm

\noindent {\bf Acknowledgments}

\noindent We thank A. Riotto and G. Dvali for useful discussions.
The work is partially supported
by the DOE under contract number DE-FG02-91ER40626.


\begin{thebibliography}{99}

\def\apj#1#2#3{Astrophys.\ J.\ {\bf #1}, #2 (#3)}
\def\ijmp#1#2#3{Int.\ J.\ Mod.\ Phys.\ {\bf #1}, #2 (#3)}
\def\mpl#1#2#3{Mod.\ Phys.\ Lett.\ {\bf #1}, #2 (#3)}
\def\nat#1#2#3{Nature\ {\bf #1}, #2 (#3)}
\def\npb#1#2#3{Nucl.\ Phys.\ {\bf B#1}, #2 (#3)}
\def\plb#1#2#3{Phys.\ Lett.\ {\bf B#1}, #2 (#3)}
\def\prd#1#2#3{Phys.\ Rev.\ {\bf D#1}, #2 (#3)}
\def\prl#1#2#3{Phys.\ Rev.\ Lett.\ {\bf #1}, #2 (#3)}
\def\prt#1#2#3{Phys.\ Rep.\ {\bf #1}, #2 (#3)}
\def\sjnp#1#2#3{Sov.\ J.\ Nucl.\ Phys.\ {\bf #1}, #2 (#3)}
\def\zp#1#2#3{Z.\ Phys.\ {\bf #1}, #2 (#3)}


\bibitem{witten} E. Witten, \npb{188}{513}{1981}.

\bibitem{unif}
U. Amaldi, W. de Boer, and H. Furstenau, \plb{260}{447}{1991};
J. Ellis, S. Kelley, and D. Nanopoulos, \plb{260}{131}{1991};
P. Langacker and M. Luo, \prd{44}{817}{1991};
C. Giunti, C. W. Kim, and U. W. Lee, Mod. Phys. Lett. {\bf A6}, 1745 (1991).

\bibitem{smbreak} K. Inoue, A. Kakuto, H. Komatsu, and S. Takeshita,
Prog. Theor. Phys. {\bf 68}, 927 (1982); {\it ibid} {\bf 70}, 330(E) (1983);
L. E. Ibanez and G. Ross, \plb{110}{215}{1982};
J. Ellis, J. S. Hagelin, D. V. Nanopoulos, and K. Tamvakis,
\plb{125}{275}{1983}.

\bibitem{nilles} H. P. Nilles, \prt{110}{1}{1984}.

\bibitem{experiment}
F. Gabbiani, E. Gabrielli, A. Masiero, and L. Silvestrini,
Nucl. Phys. {\bf B477}, 321 (1996) [hep-ph/9604387];
J. S. Hagelin, S. Kelley, and T. Tanaka, Nucl. Phys. {\bf B415}, 293 (1994).

\bibitem{gaugemedi} M. Dine and A. E. Nelson, \prd{48}{1277}{1993}
[hep-ph/9303230];
M. Dine, A. E. Nelson, and Y. Shirman, Phys. Rev. {\bf D51}, 1362
(1995) [hep-ph/9408384]; G. F. Giudice and R. Rattazzi, Phys. Rept. {\bf 322},
419 (1999) [hep-ph/9801271].

\bibitem{anomalymedi}L. Randall and R. Sundrum, Nucl. Phys. {\bf B557}, 79 (1999) [hep-th/9810155]; G. F. Giudice, M. A. Luty, H. Murayama, and R. Rattazzi,
JHEP {\bf 9812}, 027 (1998) [hep-ph/9810442].

\bibitem{esusy} A. Cohen, D. B. Kaplan, and A. E. Nelson, \plb{388}{588}{1996};
S. Dimopoulos and G. F. Giudice, Phys. Lett. {\bf B357}, 573 (1995)
[hep-ph/9507282].

\bibitem{u1a} G. Dvali and A. Pomarol, Phys. Rev. Lett. {\bf 77}, 3728 (1996)
[hep-ph/9607383];
R. N. Mohapatra and A. Riotto, Phys. Rev. {\bf D55},4262 (1997)
[hep-ph/9611273];
R.-J. Zhang, Phys. Lett. {\bf B402}, 101 (1997) [hep-ph/9702333];
A. E. Nelson and D. Wright, Phys. Rev. {\bf D56}, 1598 (1997)
[hep-ph/9702359].

\bibitem{radia} J. L. Feng, C. Kolda, and N. Polonsky, Nucl. Phys. {\bf B546},
3 (1999) [hep-ph/9810500].

\bibitem{arkani} N. Arkani-Hamed and H. Murayama, Phys. Rev. {\bf D56}, 6733 (1997)
[hep-ph/9703259]; K. Agashe and M. Graesser, Phys. Rev. {\bf D59}, 015007 (1999)
[hep-ph/9801446]; J. Hisano, K. Kurosawa, and Y. Nomura, Phys. Lett. {\bf B445},
316 (1999) [hep-ph/9810411].

\bibitem{peskin} E. A. Mirabelli and M. E. Peskin, \prd{58}{065002}{1998}
[hep-th/9712214].

\bibitem{dine} A. Anisimov, M. Dine, M. Graesser, and S. Thomas,
hep-th/0111235.

\bibitem{gherg} T. Gherghetta and A. Riotto, Nucl. Phys. {\bf B623}, 97 (2002)
[hep-th/0110022].

\bibitem{zucker1} M. Zucker, \npb{570}{267}{2000} [hep-th/9907082].

\bibitem{zucker3} M. Zucker, \prd{64}{024024}{2001} [hep-th/0009083].

\bibitem{zucker2} M. Zucker, JHEP {\bf 0008}, 016 (2000) [hep-th/9909144].

\bibitem{zuckerth} M. Zucker, ``Off-shell supergravity in five dimensions and
supersymmetric brane world scenarios'', thesis, Physikalisches Institut,
Universit${\rm \ddot{a}}$t Bonn, August 2000, BONN-IR-2000-10
[http://www.th.physik.uni-bonn.de/nilles/db/thesis
/zucker.ps].

\bibitem{kugo} T. Kugo and K. Ohasi, Prog. Theor. Phys. {\bf 104}, 835 (2000)
[hep-ph/0006231]; {\it ibid} {\bf 105}, 323 (2001) [hep-ph/0010288];
T. Fujita, T. Kugo, and K. Ohasi, Prog. Theor. Phys. {\bf 106}, 671 (2001)
[hep-th/0106051].

\bibitem{sohnius} M. F. Sohnius and P. C. West, \npb{216}{100}{1983}.

\bibitem{west} P. West, ``Introduction to supersymmetry and supergravity''
(extended second edition), Singapore, World Scientic Publishing (1990);
S. Ferrara and P. v. Nieuwenhuizen, \plb{76}{404}{1978}.

\bibitem{dewit} B. de Wit, P. G. Lauwers, and A. Van Proeyen,
\npb{255}{569}{1985}; P. Breitenlohner and M. F. Sohnius,
\npb{187}{409}{1981}.

\bibitem{onshell} L. Andrianopoli, M. Bertolini, A. Ceresole, R. D'Auria,
S. Ferrara, and P. Fr${\rm \acute{e}}$, Nucl. Phys. {\bf B476}, 397 (1996)
[hep-th/9603004]; A. Lukas, B. A. Ovrut, K. S. Stelle, and D. Waldram,
Nucl. Phys. {\bf B552}, 246 (1999) [hep-th/9806051].

\bibitem{falkowski} A. Falkowski, Z. Lalak, and S. Pokorski,
\plb{491}{172}{2000} [hep-th/0004093].

\bibitem{kyae} J. E. Kim and B. Kyae, \plb{500}{313}{2001} [hep-ph/0009043].

\bibitem{bs} Z. Chacko, M. A. Luty, A. E. Nelson,and  E. Ponton,
JHEP {\bf 0001}, 003 (2000) [hep-ph/9911323];
Z. Chacko and M. A. Luty, JHEP {\bf 0205}, 047 (2002) [hep-ph/0112172].
\end{thebibliography}
\end{document}